\documentclass[a4paper,11pt]{article}
\pdfoutput=1 
\usepackage{jheppub}
\usepackage{epsfig}
\usepackage{amsmath}
\usepackage{graphicx}

\def\A0#1{\Pi_{\rm #1}(0)}
\def\AP0#1{\Pi'_{\rm #1}(0)}
\def\be{\begin{equation}}
\def\ee{\end{equation}}
\def\bea{\begin{array}}
\def\eea{\end{array}}
\def\beqa{\begin{eqnarray}}
\def\eeqa{\end{eqnarray}}
\def\beqas{\begin{eqnarray*}}
\def\eeqas{\end{eqnarray*}}

\def\bp{\begin{picture}}
\def\ep{\end{picture}}
\def\bc{\begin{center}}
\def\ec{\end{center}}
\def\bfig{\begin{figure}}
\def\efig{\end{figure}}

\def\bit{\begin{itemize}}
\def\eit{\end{itemize}}
\def\nn{\nonumber}
\def\f{\frac}

\def\[{\left[}
\def\]{\right]}
\def\({\left(}
\def\){\right)}

\def\..{\left.}
\def\.{\right.}

\def\tm{\times}

\def\da{\dagger}

\def\al{\alpha}

\def\ep{\epsilon}

\title{Did the nHZ Gravitational Waves Signatures Observed By NANOGrav Indicate Multiple Sector SUSY Breaking?}
\author[a,b]{Xiao Kang Du}
\author[a]{Ming Xia Huang}
\author[a]{Fei Wang$^*$,}
\author[a]{Ying Kai Zhang}
\affiliation[a]{School of Physics, Zhengzhou University, Zhengzhou 450000, P. R. China}
\affiliation[b]{Institute of Physics, Henan Academy of Sciences, Zhengzhou 450046, P. R. China}
\emailAdd{feiwang@zzu.edu.cn}
\abstract{Discrete R symmetries always play an important role in low energy SUSY. The spontaneously broken of such discrete R symmetries, for example, by gaugino condensation, can lead to domain walls, which need to be either inflated away or collapse to avoid cosmic difficulties. We propose that explicitly R symmetry violation needed for collapse of domain walls can be the consequence of multiple sector SUSY breaking. The consistency constraints for the generation of non-problematic domain walls from gaugino condensation are discussed. We also study the emitted gravitational waves related to the collapse of domain walls. We find that, for SUSY breaking scale of order ${\cal O}(1){\rm GeV}$ in one of the sequestered sector (and also a low reheating temperature of order ${\rm MeV}$ if the reheating is not completed when the domain walls collapse), the peak frequency of gravitational waves emitted can lie at nHz. Such a low SUSY breaking scale can be consistency and natural in multiple sector SUSY breaking scenario. The GWs signal by NANOGrav could be a signal of such multiple sector SUSY breaking scenario and it may also indicate the existences of light goldstini at ${\rm eV}$ mass scale.  }

\begin{document}
\maketitle \indent
\newpage
\section{\label{sec-1}Introduction}
Supersymmetry (SUSY) is a theoretically well motivated framework that solves the hierarchy
problem, offers successful gauge coupling unification, provide viable dark matter candidates and naturally accommodating the observed 125 GeV Higgs etc. If SUSY is to be realized in nature, it must be spontaneously broken. In SUSY, a class of symmetries known as R-symmetries are of great interest, because existence of continuous R symmetries are crucial for spontaneous SUSY breaking for a generic Lagrangian, which follows the Nelson-Seiberg theorem~\cite{NSeiberg}. As global symmetries are believed to be violated by quantum gravitational effects, continuous R symmetries should be low energy accidents if they arise in nature. Only the discrete remnant of some gauge symmetries, the discrete R symmetry, can be exact. Discrete R symmetries could give rise to such approximate continuous R symmetries as their accidental consequence and play an important role in low energy SUSY.

Discrete R symmetries are quite common in string theory, which may arise, for example, as unbroken subgroups of higher dimensional rotation groups upon compactification.  Discrete R symmetry must be spontaneously broken to account for the small value of cosmological constant. Consequently, sheet-like topological defects, the domain walls~(DWs), will be created after the
spontaneously broken of such discrete symmetry~\cite{Kibble}. In fact, the breaking of discrete R symmetries can be quite generic in dynamic SUSY breaking models, such as from gaugino condensation.
DWs related to the discrete R symmetry breaking can form in our observable universe if the Hubble parameter during the inflation $H_I$ exceeds the R-breaking scale $\Lambda$, or alternatively if the highest temperature $T_H$ with $T_H\simeq (T_R^2 H_I M_P)^{1/4}$ after inflation exceeds $\Lambda$ (with $T_R$ the reheating temperature). The cosmological consequences of domain walls can be problematic, as their energy density soon dominates the total energy density of the universe and makes the universe extremely inhomogeneous, conflicting with the present observational results. In order that the post-inflation DWs do not cause severe problems, the DWs discrete symmetries should collapse, if such discrete symmetry is not exact. The collapse of DWs can lead to typical signals, such as gravitational waves~(GWs). We should note that such a solution can not hold for gauged discrete R symmetry although it was argued that discrete symmetries are probably gauged~\cite{WhyGaugedR}.

Direct observations of GWs is reshaping modern astrophysics and cosmology. Many important information can be extracted from direct observations of GWs because GWs
interact very weakly with matter and hence preserve almost all the features characterizing astro-physical or cosmological event~\cite{Maggiore}.
 Very recently, the NANOGrav~\cite{NANOGrav:2023gor,NANOGrav:2023hde,NANOGrav:2023hvm}, EPTA~\cite{Antoniadis:2023ott}, PPTA~\cite{Reardon:2023gzh,Reardon:2023zen,Zic:2023gta}, and CPTA~\cite{Xu:2023wog} groups published data showing
the detection of stochastic GW background in the ${\cal O}(1 \sim10)$ nHz frequency band. Many models~\cite{Kitajima:2023vre, Oikonomou:2023qfz, Addazi:2023jvg, Athron:2023mer, Yang:2023aak, Bai:2023cqj, Ellis:2023dgf, Megias:2023kiy, Franciolini:2023pbf, Li:2023yaj, Kitajima:2023cek, Fujikura:2023lkn, Vagnozzi:2023lwo, Ellis:2023tsl, Wang:2023len, Guo:2023hyp, Han:2023olf, Lambiase:2023pxd, Zu:2023olm, Shen:2023pan, Franciolini:2023wjm, Bian:2023dnv, Blasi:2023sej, Depta:2023qst, Lazarides:2023ksx, Inomata:2023zup, Cai:2023dls, Huang:2023chx, Yang:2023qlf, Broadhurst:2023tus, Wang:2023ost, Bi:2023tib, Barman:2023fad, Borah:2023sbc, Datta:2023vbs, Murai:2023gkv, Ghosh:2023aum, Gouttenoire:2023nzr, Liu:2023ymk,  Niu:2023bsr, Chowdhury:2023opo, Konoplya:2023fmh, Zhang:2023lzt, Lu:2023mcz, Xiao:2023dbb, Li:2023bxy, Anchordoqui:2023tln, Ebadi:2023xhq, Liang:2023fdf, Abe:2023yrw, Unal:2023srk, Figueroa:2023zhu, Yi:2023mbm, Chiang:2020aui} had been proposed to explain this signal.
The  Gravitational Waves (GWs) generated by the collapse of DWs can possibly lie in such frequency band and be detected by these GW experiments.

To protect the VEV of the superpotential and account for the small cosmological constant, the breaking of the discrete R symmetry and SUSY breaking are closed tied together. The corresponding SUSY breaking can occur in a single sector or in multiple sectors, as string theoretic constructions routinely predict a multiplicity of geographically sequestered sectors and any number of them could independently break SUSY.
So, the related GWs signals from DWs by discrete R symmetry breaking could possibly carry information related to the SUSY breaking sector.

This paper is organized as follows. In Sec~\ref{sec-2}, we show that the explicit discrete R violation can naturally be induced in multiple sector SUSY breaking scenario. In Sec~\ref{sec-3}, the consistency constraints for the generation of non-problematic domain walls from gaugino condensation are discussed.  In Sec~\ref{sec-4}, the GWs signals related to DWs collapse are discussed. Sec~\ref{sec:conclusions} contains our conclusions.

\section{\label{sec-2}Induced Explicitly Discrete R Symmetry Violation For A Sequestered Sector}
It is well known that SUSY must be spontaneously broken if it is to be realized in nature.
While single sector SUSY breaking is convenient, the possibility of independently SUSY breaking from sequestered sectors is well-motivated from top-down considerations.

As noted in~\cite{1002.1967}, in the limit in which these sequestered sectors are completely decoupled, even gravitationally, they enjoy an N-fold enhanced Poincare
symmetry because energy and momentum are separately
conserved within each sector. Likewise, SUSY is similarly enhanced because it is the maximal extension of space time symmetry. The $SUSY^N$ enhancement is an accidental consequence
of the decoupling limit, the gravitational interactions break explicitly the $SUSY^N$ down to a diagonal combination corresponding to the genuine supergravity (SUGRA) symmetry. The "orthogonal" $SUSY^{N-1}$ are only approximate global symmetries. The same holds for their corresponding R symmetries, including their discrete version $Z_R^N\equiv \otimes\prod\limits_{i}^N (Z_{M_i})$.

When the F-term breaking occurs independently in each sector~\cite{1002.1967}, the spontaneously broken scales $F_i$ satisfy that $F_i \equiv r_i F_{eff}$ ($\sum_i r_i^2 = 1$), yielding a corresponding goldstino for that sector. One linear combination of them will be eaten by the gravitino via the super-Higgs mechanism, the remaining goldstini fields are still propagating degree of freedoms. Different SUSY breaking sectors themselves still interact only through SUGRA.
When two or more SUSY breaking sectors have direct couplings to the SSM to mediate SUSY breaking, for example, by gauge mediation mechanism, induced couplings between each sequestered sector can arise.
 Each sequestered SUSY breaking sector is assumed to couple to the SSM sector via an operator suppressed by $\Lambda_i$. Other than spontaneously breaking of discrete R symmetry in each sequestered SUSY breaking sector, new induced discrete $Z_{M_i}$ breaking effects for sequestered sector $i$ can be generated only through the SSM sector. The leading discrete R-breaking transmitting couplings between a SUSY breaking sector and the SSM sector are given by the gaugino-mass and A-term operators
 in the sequestered sectors~\cite{1002.1967}
\beqa
\int d^2\theta \f{X_i}{\Lambda_i} W^a W^a,~~\int d^2\theta \f{X_i}{\Lambda_i} \Phi^\da\Phi~,
\eeqa
with the F-term SUSY breaking $F_i\neq 0$ for the $i$ sector characterized by a chiral superfields $X_i$ with its Vacuum Expectation Value (VEV) $\langle X_i\rangle=M_i+\theta^2 F_i$. The $W^a$, $\Phi$ represent the gauge field strengths and chiral superfields of the SSM, respectively. The sfermion mass operators
\beqas
\int d^4\theta \f{X_i^\da X_i}{\Lambda_i^2} \Phi^\da\Phi~,
\eeqas
do not provide necessary R-breaking transmission, unless $X_i$ has a lowest component VEV to give effectively A-term operators.

Similar to the continuous $U(1)_{R}$ case, the induced couplings between sectors $i$ and $j$ can be generated by the Feynman diagrams involving the discrete R-breaking transmitting couplings and can be parameterized as
\beqa
{\cal O}_{\not{R}_i \not{R}_j}&\sim& \(\f{1}{16\pi^2}\)^{n_{ij}}\int d^4\theta X_i^\da X_j~,\nn\\
{\cal O}_{\not{R}_i}&\sim& \(\f{1}{16\pi^2}\)^{n_{i}}\f{1}{\max\{\Lambda_i,\Lambda_j\}}\int d^4\theta X_i X_j^\da X_j+h.c.~,\nn\\
{\cal O}_{\not{R}_j}&\sim& \(\f{1}{16\pi^2}\)^{n_{j}}\f{1}{\max\{\Lambda_i,\Lambda_j\}}\int d^4\theta X_j X_i^\da X_i+h.c.~,
\label{R-break}
\eeqa
with $n_{ij},n_i,n_j\geq 1$ for the loop factors, since there is always at least one loop of SSM fields involved in the diagram (for example, $n_{ij} = 3$ for gauge mediation). Such loop suppressed explicit discrete R symmetry violating terms can be useful to induce the annihilation of the domain walls that produced from spontaneously discrete R symmetry breaking.

\section{\label{sec-3}Domain Walls from Gaugino Condensation In Some Sequestered Sector}
 To achieve an almost vanishing cosmological constant, the VEV of the superpotential $\langle W_0\rangle $ is required to be small to cancel the the low-scale SUSY-breaking contribution $|F|^{2}$~\cite{1506.00426}. Thus, it is inevitable to invoke the discrete R-symmetry to prevent $W_0$ from being very large, with the non-vanishing small VEV of the superpotential generated by the spontaneous breaking of such a discrete R-symmetry. Spontaneous breaking of discrete R symmetries can be triggered in pure supersymmetric gauge theories by gaugino condensation and also by the squark condensation in theories with massive matter fields~\cite{Intriligator:2007cp}.

 The disappearance of domain wall related to discrete R symmetry breaking after inflation might be
explained by explicit breaking of the symmetry. This can be accomplished if there is a large enough explicit R breaking and such breaking is plausible~\cite{1005.3613}. The effects of the explicit R-symmetry breaking can be parameterized through a constant $w_c$ in the superpotential with the spontaneously breaking of discrete R symmetry through gaugino condensation $W_0\supseteq \Lambda^3 \al^k$~\cite{0802.4335}. Domain walls are produced at the phase transition associated with the scale $\Lambda$, and they will annihilate as a result of the explicit breaking $w_c$.


We propose to adopt such setting of~\cite{0802.4335} for sector $i$ in multiple sector SUSY breaking scenario. The superpotential in certain sequestered sector $i$ takes the gaugino condensation form
\beqa
W\supseteq  N\Lambda^3 \al^k+w_c~,
\eeqa
from pure SUSY SU(N) Yang-Mills theory with $\al$ the $N$-th root of unity. The gaugino
condensation spontaneously breaks the $Z_{2N}$ from $U(1)_R$ by instanton breaking further down to $Z_2$, leading to DWs at epochs of the condensations in the early universe. The discrete $Z_{2N}$ symmetry is also violated explicitly from the $w_c$ term in the superpotential, which can arise from the induced discrete R symmetry breaking operator in~(\ref{R-break}) as
\beqa
w_c\simeq c_{{\cal O}_{\not{R}_i \not{R}_j}} \sum\limits_{j \neq i}\(\f{1}{16\pi^2}\)^{n_{ij}} \( M_i F_j+ F_i M_j\)+c_{{\cal O}_{\not{R}_i}}\(\f{1}{16\pi^2}\)^{n_{i}}\sum\limits_{j\neq i}\f{2 (M_i F_j+M_jF_i) }{\max\{\Lambda_i,\Lambda_j\}} ~.
\eeqa
 In the case in which the SUSY breaking in the $i$ sector does not dominant the total SUSY breaking, we expect that
 \beqa w_c\simeq c_0 M_i \sum\limits_{j\neq i}F_j\sim c_0 M_i F_{eff}~,\eeqa
 with $F_{eff}=\sqrt{\sum\limits_{i}F_i^2}$ and $c_0\ll 1$ some combinations of small loop factors, assuming all $M_i$ at same order.

The properties of domain walls can be characterized by two model-dependent quantities, the tension $\sigma$ and the thickness $\delta$. In general, the wall thickness is roughly given by Compton wavelength of the field which causes the spontaneous breaking of
discrete symmetry, while the tension is estimated in terms of the height of the potential energy that
separate the degenerate minima.
 The surface energy density of domain walls takes the form~\cite{dvali,hep-ph:9804409}
\beqa
\sigma\simeq 2N \Lambda^3 |1-e^{\f{2\pi i}{N}}|\simeq  4\pi \Lambda^3~.
\eeqa
From the scalar potential in SUGRA for F-term breaking
 \beqa
 V_F &=& e^\f{K}{M_P^2}\[K_{ij^*}^{-1} \(W^{i}+\f{K^i}{M_P^2}W\)\(W^{j^*}+\f{K_{j^*}}{M_P^2}W^*\)-3 \f{|W|^2}{M_P^2}\]\nn\\
 &\simeq & |F_i|^2-3\f{N^2|\Lambda^3|^2}{M_P^2}-3 N\Lambda^3 (\al^k+\al^{-k}) w_c/M_P^2~,
 \eeqa
we can obtain the potential values at the distinct $N$ minimum and the typical size of the energy difference between the adjacent states
\beqa
\epsilon\simeq  A_0 \Lambda^3 w_c/M_P^2~,
\eeqa
for some constant $A_0\sim {\cal O}(10)$.

 The close tie between the scale of SUSY breaking and the scale of spontaneously R symmetry breaking to guarantee the smallness of cosmological constant can be spoiled in multiple sector SUSY breaking scenarios. Note that the vacuum energy in the sequestered sector need not be canceled to a tiny value as the nearly cancelation of cosmological constant should be hold after taking into account all the sequestered sectors.
  The broken scale $\Lambda$ of discrete R symmetry via gaugino condensation in sector $i$ needs not related to the the corresponding F-term breaking parameter $F_i$ as $s\equiv |\f{N\Lambda^3}{\sqrt{3}F_i M_P}|=1$. The choice $s\gg1$ or $s\ll1$ can also be allowed.

The gaugino condensation scale thus satisfies
\beqas
N \Lambda^3\simeq s F_i M_P\lesssim  s F_{eff}M_P=s m_{3/2} M_P^2~,
\eeqas
because in multiple sector SUSY breaking $F_i\lesssim F_{eff}$ for $m_{3/2}=F_{eff}/M_P$.
Therefore, it is always reasonable to parameterize $\Lambda^3=b m_{3/2} M_P^2$ with $b\ll1$, either with $s\ll 1$, $F_i\ll F_{eff}$ or large $N$. The $w_c$ term in the superpotential that explicitly break the discrete symmetry can be parameterized as
\beqas
w_c\simeq c_0 F_{eff} M_i= c_0 m_{3/2} M_P M_i=c_0\f{M}{M_P} m_{3/2} M_P^2~.
\eeqas
So, $w_c$ can be parameterized as $w_c\equiv a m_{3/2} M_P^2$ with $a\ll1$. The factor $a$ origins from loop diagrams, so it can naturally be small because the discrete R violation operators are multiple-loop suppressed, for example, by at least 3-loop in gauge mediation.

 After the DWs are formed, they quickly follow a scaling solution~\cite{scalinglaw} such that there is about one domain wall per Hubble horizon. Assuming that the evolution of the domain walls is indeed given by the scaling law, the energy density of the domain walls is
\beqa
\rho_{DW}\simeq \f{\cal A}{t} \sigma \sim {\cal A} \sigma H~,
\eeqa
 with the area parameter ${\cal A}\sim {\cal O}(1)$ taking an almost constant value during the scaling regime.
 The domain walls can collide with one another and disappear when the tension and the pressure on the wall due to the bias (that is, the explicit violation of the discrete symmetry) become comparable~\cite{Vilenkin}. The Hubble parameter at the decay time can be estimated from
 $\ddot{x}_i\simeq {\epsilon}/{\sigma}$ to give $H\sim {\epsilon}/{\sigma}$,
 which should be earlier than $H\sim \sigma/M_P^2$, otherwise DWs will come to dominate the energy density of the universe and drive the universe to large inhomogeneous and anisotropic. This constraint gives a lower bound on the bias:
\beqa
\epsilon \gtrsim \sigma^2/M_P^2.
\label{bias:constraint}
\eeqa

However, it was argued in~\cite{1005.3613} that the picture in which domain walls form with characteristic scale $\Lambda^3 \simeq m_{3/2}M_P^2$ and annihilate due to some smaller explicit R breaking effect is not viable. Besides,it was also noted in that paper that the interactions responsible for the explicit R breaking must make the dominant contribution to the superpotential.

We revisit the strong constraints on the walls annihilation before gravitational collapse that exclude the simplest gaugino condensation plus explicit R breaking scenario.
To evade its gravitationally instability and collapsing to form singularities, the gravitational Schwarzchild radius for a sphere of radius $R$ centered on the wall, which is given by $r_*\sim G \pi R^2\sigma$, should be less than the radius $R$. This requirement gives $R<(2\pi G \sigma)^{-1}$~\cite{Vilenkin}.  Therefore, the Schwarschild radius associated with the DWs tension
in a horizon should be smaller than the horizon of the DWs, which leads to $H\gg (2\pi G \sigma)$. Therefore, the following constraints
\beqa
H\sim {\epsilon}/{\sigma}~,~~H\gg (2\pi G \sigma)~,~
\label{DW:consistency}
\eeqa
should be compatible. The two conditions in~(\ref{DW:consistency}) reduce to
\beqa
H\sim a m_{3/2},~~H \gtrsim b {m_{3/2}}~,
\eeqa
in our case with $\sigma\sim 4\pi b m_{3/2} M_P^2$ and $\epsilon\sim A_0 a b m_{3/2}^2 M_P^2$.
They can be mutually compatible with each other and at the same time also consistent with the constraints in~(\ref{bias:constraint}), which gives $a\gtrsim b$. Such a condition agrees with the conclusion in~\cite{1005.3613} that the interactions responsible for the explicit R breaking must make the dominant contribution to the superpotential. Obviously, DWs cannot be created from the beginning if the energy differences between degenerate vacuums are sufficiently large (that is, when the bias is large). Constraint on the bias from the prediction of percolation theory gives~\cite{1703.02576}
 \beqa
 \f{V_{bias}}{V_0}<\ln\( \f{1-p_c}{p_c}\)=0.795~,
 \eeqa
 with $p_c=0.311$~\cite{percolation} and $V_0$ being the height of the potential barrier between the adjacent minima.  Large scale DWs are expected to be formed as long as the above condition
is satisfied. In our case, we can estimate that
 \beqa
 \f{V_{bias}}{V_0}&\approx& 2\f{a}{b}\[\cos(\f{2k\pi}{N})-\cos(\f{2(k+1)\pi}{N})\]\nn\\
   &\approx& 2\f{a}{b}\f{\pi}{N}\sin\[\f{2\pi}{N}(k+1/2)\]\nn\\
   &<& \f{a}{b}\f{2\pi}{N}<0.795~,
 \eeqa
 will lead to $a/b<0.8 N/2\pi$. So, even for $a \gtrsim b$, the formation of large scale DWs can still be allowed when the value of $N$ is not very small, for example, $N>7$.
\section{\label{sec-4} Gravitational Waves With the Collapse of DWs}
The energy stored inside the DWs will be released into light degrees of freedom and GWs. In fact, the collisions of the DWs can deviated from the spherical symmetry so as that the emission of GWs are possible. The GWs emitted are expected to peak at the frequency corresponding to the horizon scale at the decay time~\cite{0802.4335} because the DW network follows the scaling solution and the curvature radius is the Hubble horizon.
The density parameter of the GWs is defined by
\beqa
\Omega_{gw}(f)\equiv\f{1}{\rho_c}\f{d \rho_{GW}}{d\log f}~,
\eeqa
 with $\rho_{GW}$ the energy density of the GWs, $\rho_c$ the critical energy density and $f$ the
frequency of GWs.
The present density parameter $\Omega_{gw}^0$  and the frequency of the GWs $f_0$ are given by
\beqa
\Omega_{gw}^0=\Omega_{gw}^*\(\f{a_*}{a_0}\)^4\(\f{H_*}{H_0}\)^2~,~~f_0=f_*\(\f{a_*}{a_0}\),
\eeqa
due to red-shift by the cosmic expansion. The quantity with $'*'$ are the value at the formation of the GWs. In our case, the DWs decay at $H_*\simeq a m_{3/2}$ with $a\ll 1$.

 The peak frequency tody is estimated by
\beqa
f_0&=&\f{a(t_*)}{a(t_0)} H_*\f{f_*}{H_*}~,\nn\\
&\simeq& 9.3\tm 10^{-14}\(\f{61}{g_{*s}(T*)}\)^{1/3}\(\f{\rm 1 GeV}{T^*}\) H_*~,\nn\\
&\sim &  10^{-9} \(\f{61}{g_{*s}(T*)}\)^{1/6} \(\f{\rm 1 GeV}{T^*}\) \(\f{a m_{3/2}}{10^{-9}{\rm GeV}}\){\rm Hz}~,
\eeqa
with $g_{*}$ and $g_{*s}$ the effective relativistic degrees of freedom  for the energy density and the entropy density, respectively. The temperature can be transformed into the Hubble parameter via the formula
\beqa
H=\f{1}{2t}=\sqrt{\f{8\pi^3}{90}g_*}\f{T^2}{M_P}~,
\eeqa
in radiation dominant era.
We can see that for $H*\sim a m_{3/2}=10^{-9} {\rm eV}$, the peak frequency of the GWs emitted lies at $10^{-9} {\rm Hz}$. The corresponding temperature is given by
\beqas
T_*\approx (am_{3/2}M_P)^{1/2}\sqrt{\f{90}{8\pi^3g_*}}\sim {\cal O}(1)~{\rm GeV}~,
\eeqas
which is also of order the F-term breaking scale $F_i$ in the $i$ sector. Such a low $F_i$ is natural in multiple sector SUSY breaking scenarios, which allows $F_i\ll F_{eff}$. The factor $b$, which characterizes the explicitly discrete R violation and origins from multiple loops, naturally gives a suppressed factor of order $10^{-6}\sim 10^{-9}$ (for example, from at least three loops for GMSB). The factor $a$, which should satisfies $b\gtrsim a$, should also be very small. As noted previously, such a small $a$ is natural for $s\ll1$, $F_i\ll F_{eff}$ and large $N$.

According to the quadrupole formula, the power of the gravitational radiation is given by $P\sim G
\dddot{Q}_{ij} \dddot{Q}_{ij}\sim M_{Wall}/t^2$ with $M_{wall}=\sigma {\cal A} t^2$ the energy of DWs.
 The value defined by
\beqa
\tilde{\epsilon}_{gw}\equiv \f{1}{G {\cal A}^2\sigma^2}\(\f{d\rho}{d\log f}\)_{peak}~,
\eeqa
is shown by numerical results to almost keep as a constant $\tilde{\epsilon}\simeq 0.7\pm 0.4$ after DWs enter into the scaling regime. The peak amplitude at the annihilation time of domain walls is given as~\cite{1703.02576}
\beqa
\Omega_{gw}(H_*)_{peak}=\f{1}{\rho_c(H_*)}\(\f{d \rho_{gw}(H_*)}{d\log f}\)_{peak}=\f{8\pi G^2{\cal A}^2\sigma^2\tilde{\epsilon}_{gw}}{2H_*^2}~.
\eeqa
Here the production of GWs is assumed to suddenly terminate at $H=H_*$ and happen during the radiation dominated era. The present density parameter $\Omega_{gw}^0$ can be estimated as~\cite{1703.02576}
\beqa
\Omega_{gw}(t_0) h^2&=&\Omega_{rad} h^2\(\f{g_*(T_{ann})}{g_{*0}}\)\(\f{g_{*s0}}{g_{*s}(T_{ann})}\)^{4/3}\Omega_{gw}(t_{ann}),~\nn\\
&\sim& 1.16\tm 10^{-5}{\cal A}^2\(\f{b}{a}\)^2,\nn\\
&\sim & 1.16\tm 10^{-5}~,
\eeqa
for $a/b\gtrsim 1$, with $\Omega_{rad} h^2= 4.15\tm 10^{-5} $ the density parameter of radiations at the present time. So, we conclude that, in our case the GWs with $f^0_{peak}\sim 10^{-9}$ {\rm Hz} and $\Omega_{gw}(t_0) h^2\sim 1.16\tm 10^{-5}$ can be the consequence of DW collapse. Such a region lie in the sensitivity of recent NANOGrav GWs experiment.

The previous discussions based on the assumption that the reheating of the universe is completed before the collapse of the DWs so as that the universe is radiation-dominated afterwards until the matter-radiation equality. It is possible that the reheating of the universe is not completed before the collapse of the DWs. Then the universe is matter-dominated until the reheating. Therefore, the GWs are diluted by the entropy production of the inflaton. This possibility is pointed out in~\cite{0802.4335}. The corresponding red-shift, thus, the peak frequency today $f_0$, depends on the reheating temperature $T_R$. We have for the peak frequency today $f_0$
\beqa
f_0&=&\f{a(t_*)}{a(t_0)}f_*
=\f{a(t_*)}{a(t_0)} H_*\f{f_*}{H_*},\nn\\
&\simeq& 2\tm 10^{-8}\(\f{T_R}{1 MeV}\)^{1/3}\(\f{a m_{3/2}}{10^{-9}{\rm eV}}\)^{1/3}{\rm Hz},
\eeqa
where the scale factor can be estimated to be
\beqa
\f{a_*}{a_0}&=&\f{a_*}{a_R}\f{a_R}{a_0},\nn\\
&\simeq& \(\f{\pi^2 g_*}{30}\)^{1/3} T_{R}^{4/3}(H_*M_P)^{-2/3}
\(\f{g_{*s}(0)}{g_{*s}(T_R)}\)^{1/3}\f{T_0}{T_R},\nn\\
&\simeq&\(\f{\pi^2 g_*}{30}\)^{1/3}\(\f{g_{*s}(0)}{g_{*s}(T_R)}\)^{1/3}M_P^{-2/3} a^{-2/3} m_{3/2}^{-2/3}T_R^{1/3}.
\eeqa
We can see that the peak frequency of GWs can lie at ${\rm nHz}$ unless the reheating temperature is very low, of order MeV.

The density parameter $\Omega_{gw}^0$ should changed accordingly as the energy density of the universe is dominated by that of the inflaton, which behaves as non-relativistic matter, when domain walls are annihilated. We have
\beqa
\Omega_{gw} h^2(t_0)=\Omega_{rad} h^2\(\f{g_*(T_{reh})}{g_{*0}}\)\(\f{g_{*s0}}{g_{*s}(T_{reh})}\)^{4/3}\(\f{H_{reh}}{H_{ann}}\)^{2/3}\Omega_{gw}(t_{ann}),
\eeqa
with $H_*^2 a_*^3= H_{reh}^2 a_{reh}^2$ and $\rho_* a_*^3=\rho_{reh} a_{reh}^3$ at the matter-dominated region. The reheating temperature is given by~\cite{hep-ph:0005123} to be
\beqa
H_{reh}=\[\f{5\pi^2 g_*(T_{reh})}{72 }\]^{1/2}\f{T_{reh}^2}{M_P}.
\eeqa
So, we can obtain
\beqa
\Omega_{gw}h^2(t_0)&\simeq& 1.16\tm 10^{-5} \(\[\f{5\pi^2 g_*(T_{reh})}{72 }\]\)^{1/3}\(\f{T_{reh}^2}{ a m_{3/2}M_P}\)^{2/3}~,\nn\\
&\sim& 1.02\tm 10^{-9}\(\f{T_{reh}}{1~{\rm MeV}}\)^{4/3}\(\f{10^{-9} {\rm eV}}{a m_{3/2}}\)^{2/3}\(\f{b}{a}\)^2.
\eeqa
We can see that, for $a m_{3/2}=10^{-9} {\rm eV}$ and $T_R= 1 {\rm MeV}$, the energy density $\Omega_{gw}h^2$ and $f_{peak}^0$ can be respectively $1.02\tm 10^{-9}$ and $2\tm 10^{-8}{\rm Hz}$, which lies just in the new signal region by NANOGrav. Again, the value $a m_{3/2}=10^{-9} {\rm eV}$ indicates that the F-term breaking scale $F_i$ in the $i$ sector should lie at ${\cal O}(1){\rm GeV}^2$ order. We should note that the peak frequency $f_0$ and density parameter $\Omega_{gw}^0$  depend only on the combination $am_{3/2}$, therefore, only the corresponding $F_i$ value matters.
The SUSY breaking in $i$ sector by low scale $F_i\sim {\cal O}(1){\rm GeV}^2$ seems to indicate a very light goldstini. However, as discussed in~\cite{1002.1967}, the corresponding goldstini in multiple sector SUSY breaking acquire tree-level masses $2m_{3/2}\simeq 2 F_{eff}/M_P$ instead of the naively expected value $F_i/M_P$, which can be the consequence of anomaly mediation SUSY breaking. The SSM fields can couple more strongly to the goldstini than to the
gravitino. The GWs signal by NANOGrav could indicate the existences of light goldstini at {\rm eV} mass scale.
\section{\label{sec:conclusions} Conclusions}
  Discrete R symmetries always play an important role in low energy SUSY. The spontaneously broken of such discrete R symmetries, for example, by gaugino condensation, can lead to domain walls, which need to be either inflated away or collapse to avoid cosmic difficulties. We propose that explicitly R symmetry violation needed for collapse can be the consequence of multiple sector SUSY breaking. Such a setting is fairly economical without the needs of additional assumptions on the explicitly breaking scale of discrete R symmetry.
  The consistency constraints for the generation of non-problematic domain walls from gaugino condensation are discussed. We also study the emitted gravitational waves related to the collapse of domain walls. We find that, for SUSY breaking scale of order ${\cal O}(1)$ {\rm GeV} in one of the sequestered sector (and also a low reheating temperature of order MeV if the reheating is not completed when the domain walls collapse),  the peak frequency of gravitational waves emitted can lie at nHz. Such a low SUSY breaking scale can be consistency and natural in multiple sector SUSY breaking scenario. The GWs signal by NANOgrav could be a signal of such multiple sector SUSY breaking scenario and it may also indicate the existences of light goldstini at ${\rm eV}$ mass scale.

  We should note that again the peak frequency $f_0$ and density parameter $\Omega_{gw}^0$  depend only on the combination $am_{3/2}$ (and the reheating temperature $T_R$). Therefore, only the corresponding scale of $F_i$ matters.

\begin{acknowledgments}
 This work was supported by the
Natural Science Foundation of China under grant numbers 12075213; by the Key Research Project of Henan Education Department for colleges and universities under grant number 21A140025.
\end{acknowledgments}

\end{document}